\documentclass[conference]{IEEEtran}
\IEEEoverridecommandlockouts
\usepackage{amsmath,amssymb,amsfonts}
\usepackage{graphicx}
\usepackage{textcomp}
\usepackage{xcolor}
\usepackage[]{graphicx}
\usepackage{algorithm}
\usepackage{comment}
\usepackage{algpseudocode}
\usepackage{pgfplots}
\usepackage[skip=0pt]{caption}
\usepackage[]{amssymb}
\usepackage{amsmath}
\usepackage{amsthm}

\newtheorem*{namedtheorem}{Property 1}
\usepackage{float}
\usepackage{amsmath}
\usepackage{mathtools}
\usepackage{slashbox}
\usepackage{multirow}
\usepackage{subfigure}
\graphicspath{{images/}}
\def\BibTeX{{\rm B\kern-.05em{\sc i\kern-.025em b}\kern-.08em
    T\kern-.1667em\lower.7ex\hbox{E}\kern-.125emX}}
\begin{document}

\title{Secured Traffic Monitoring in VANET\\
}

\author{\IEEEauthorblockN{Ayan Roy}
\IEEEauthorblockA{\textit{Department of Computer Science} \\
\textit{Missouri University of Science and Technology, USA}\\
%Rolla, MO 65409 , USA \\
ar3g3@mst.edu}
\and
\IEEEauthorblockN{Sanjay Madria}
\IEEEauthorblockA{\textit{Department of Computer Science} \\
\textit{Missouri University of Science and Technology, USA}\\
%Rolla, MO 65409 , USA \\
madrias@mst.edu}

}

\maketitle
\begin{abstract}
  Vehicular Ad hoc Networks (VANETs) facilitate vehicles to wirelessly communicate with neighboring vehicles as well as with roadside units (RSUs). However, an attacker can inject inaccurate information within the network that can cause various security and privacy threats, and also disrupt the normal functioning of any traffic monitoring system. Thus, we propose an edge cloud based privacy preserving  secured decision making model that employs a heuristic based on vehicular data such as GPS location and velocity to authenticate traffic-related information from the ROI under different traffic scenarios. The effectiveness of the proposed model has been validated using VENTOS, SUMO, and Omnet++ simulators, and also, by using a simulated cloud environment. We compare our proposed model to the existing state-of-the-art models under different attack scenarios. We show that our model is effective and capable of filtering data from malicious vehicles, and provide accurate traffic information under the influence of at least one non-malicious vehicle.
\end{abstract}


\section{Introduction}
Vehicular Ad hoc Networks (VANETs) allow wireless communication from vehicle-to-vehicle (V2V), and vehicle-to-infrastructure (V2I) such as with road side units (RSU) for better traffic management. %\cite{yao2017using}.
Using the dedicated short range communication (DSRC) protocol, every vehicle broadcasts information about traffic events such as accidents, traffic congestion, and traffic violations to nearby vehicles as well as road side infrastructures. However, the presence of malicious vehicles at the region of interest (ROI) can negatively influence the traffic monitoring of the region. Bluetooth based traffic monitoring systems such as Clearview Intelligence's M830 leverage the unique MAC address of Bluetooth devices inside the car and their entry, exit times form the zones to determine the traffic flow. However, the malicious vehicles can perform Denial of Service attack by switching off their devices, on the event of which the device records fewer vehicles within the region. Also, a naive way of thinking to solve the problem is to collect location and velocity information from vehicles and discard inconsistent information with the RSU's location and speed. However, such a solution will not be effective where the malicious vehicles send incorrect velocity information to the RSU using V2I communication. A comprehensive review of the solutions to preserve the privacy, authenticity and security of the messages disseminated in VANET has been provided in \cite{manivannan2020secure}. Existing strategies such as the peer authentication model \cite{yang2018blockchain}, threshold based or majority voting \cite{shao2015threshold}, and the reputation-based system \cite{li2012reputation} provide traffic monitoring by assuming the concept of adversarial parsimony \cite{golle2004detecting} but it becomes challenging to validate the vehicular responses when the malicious vehicles are in the majority. The majority of the vehicles can be compromised when a group of malicious vehicles forming the road network performs a collusion attack on the system. Also, in the case of a small network, an attack can be propagated among the majority of vehicles when a compromised vehicle communicates using V2V communication to its nearby vehicles, thus, compromising them.  

%As a motivating example, an inaccurate traffic response originating from the ROI can detour an emergency vehicle from taking the shortest path for evacuation in case of a disaster.Furthermore,
An increase in the autonomy of vehicles makes the infrastructure even more vulnerable to security attacks. The remote hacking of a Jeep Cherokee in 2015 \cite{greenberg2015hackers} highlights the feasibility of such attacks. % which can be extended to majority vehicles deployed at the ROI. %Also, the GPS location value reported by a malicious vehicle can be manipulated to disrupt any traffic monitoring system. 
Additionally, the breach in the privacy of individual vehicle information can lead to %dire consequences, such as 
unauthorized tracking of officials as well as vehicles identity theft. The purpose of an attacker is to profile a driver's habits based on GPS location, velocity, acceleration, and the unique ID of the vehicle. Under such constraints, it is necessary to ensure the anonymity of the vehicles and unlinkability of the shared information to their originating vehicles. Anonymity ensures that every vehicle remains anonymous while exchanging information whereas, unlinkability ensures the inability to trace the identity of a vehicle based on the information exchanged by it. Furthermore, such a system should ensure conditional privacy, which means that the identity of the attacker can be revealed in case of a conflict.

%To resolve the issue, 
The United States Department of Transport (USDOT) introduced Security Credential Management System (SCMS) \cite{whyte2013security} that leverages V2V and V2I communication among vehicles, and public key infrastructure (PKI) to ensure message integrity, authenticity, privacy, and interoperability. Due to the lack of misbehavior detection algorithm, if authentic vehicles misbehave and provide inaccurate traffic-related information from the ROI, there is no algorithm in place to validate the information and filter the malicious vehicles and their responses from the network, and thus, it can obtain inaccurate traffic information. This problem is elevated even more if the malicious vehicles at the ROI form the majority and try to disrupt the traffic monitoring system. To address these shortcomings, we design a global misbehavior detection algorithm which is yet to be defined by USDOT. We propose a secure and privacy preserving decision making model by leveraging the PKI and an edge cloud based infrastructure to validate traffic-related information from the ROI and filter malicious vehicles, even if they are in majority (like under collaborative or DDoS attack), within the ROI. Each edge server is associated with a different region and is connected to a centralized server. By leveraging the concept of the edge server, the decision making model is brought closer to the concerned ROI, which reduces the latency in the decision and reduces the bottleneck on the centralized server. The main contributions of the proposed work are as follows:
%Our aim is to design and develop a heuristic by leveraging the reported vehicular information such as GPS location, velocity and authentic vehicle ID to validate the traffic-related information from the ROI by preserving the privacy of information of the individual vehicles from other nearby vehicles. The main contributions of the proposed work are as follows:
\begin{itemize}
    \item Develop a privacy preserving and secure heuristic based solution that overcomes the shortcomings of the current state-of-the-art models and validates the traffic-related information from the ROI using the recorded GPS location of each vehicle, the vehicles' velocity and encrypted neighboring vehicles' IDs under the influence of at least one non-malicious vehicle within the ROI.  It is unlike the assumption of the majority of non-malicious vehicles considered in other state-of-the-art models. We also consider an event recorded by an individual vehicle because the velocity of the vehicle may not always reflect the event at that ROI (such as when a vehicle is moving with a low velocity along the service lane in a non-congested road).  
    \item Design a %new 
    dynamic data structure called the \emph{Decision Similarity Graph} based on the vehicle location, and leverage the \emph{Point of Conflict} concept to filter malicious vehicles within the ROI using the conflicting event recorded by any two neighboring vehicles.
    \item Validate the effectiveness of the proposed model as compared to the other existing state-of-the-art models under different scenarios using the simulations, and show that the proposed model effectively validates the traffic-related information and filters the malicious vehicles and their responses from the network. The proposed model has been compared against the state-of-the-art models that leverage the V2X communication infrastructure to validate the traffic-related information. %from the ROI.
\end{itemize}

\section{RELATED WORK}
%To design secure VANETs, %many studies focus on the privacy and security of the system. researchers have proposed threshold based authentication systems where a threshold for every vehicle decides whether data is acceptable or not. 
\section{Related Work}
Here, we review some of the existing models and highlight their merits and limitations under different scenarios. %We categorize the models based on different techniques used for data credibility assessment. 
\subsection{Majority Voting Model or Threshold Based Model}
A weighted majority voting based model has been proposed in \cite{huang2014social} where the majority of the vehicles are considered reliable and the message generated by a vehicle closer to the event has a higher weight than the vehicle at a distance. However, the proposed model may result in an inaccurate data reporting if the vehicle closer to the event is malicious or majority of the vehicles are malicious. 
The authors in \cite{taie2017novel}\cite{shao2015threshold} have proposed threshold based data authentication schemes in which a vehicle considers a message as credible if it has been authenticated by a threshold number of vehicles. However, the proposed schemes are highly vulnerable to ballot stuffing, bad mouthing, collusion, sybil attack and in situation when the number of malicious or compromised vehicles authenticating an incorrect message is greater than or equal to the threshold. A privacy-preserving traffic monitoring system is proposed in \cite{zhu2019traffic} where the vehicles share their speed information with the nearby vehicles. To preserve the privacy, the speed information is perturbed with noise and the security of the information is ensured using homomorphic cryptosystem. The traffic scenario of the region is decided by computing the average of the speed information shared by all the vehicles in the region. However, the proposed model inherently assumes that the vehicle does not tamper its speed information to disrupt the traffic scenario of the region. 
\subsection{Reputation Based Model}
The authors in \cite{li2012reputation} have proposed a reputation based announcement scheme in which the credibility of the message generated by a vehicle depends on the reputation score of the vehicle which is obtained by the reliability of its broadcast messages in the past. However, the model is susceptible to on-off attack, where a vehicle with a high prior reputation score performs maliciously or is compromised by an attacker.
\subsection{Peer Authentication Model}
A blockchain-based reputation system has been proposed in \cite{yang2018blockchain} in which the credibility of the messages generated by a vehicle is determined by the prior reputation of the vehicle and by ratings from the nearby vehicles. The authors in \cite{huang2017distributed} has also proposed a distributed reputation management system in which the credibility of the broadcast message originating from a vehicle is determined by the ratings obtained from its one-hop neighbors. The authors in \cite{hasrouny2018trust} have also proposed a peer authentication based trust management model based on trust scores. However, the proposed models are highly vulnerable to ballot stuffing and bad-mouthing attack, in which the ratings of an individual vehicle can be influenced by nearby vehicles. The proposed model also provides incorrect traffic information when the majority of the vehicles are malicious or are compromised by an attacker.

Our proposed model differs from the above mentioned models/schemes in the following aspects: 1) involves no peer authentication, majority voting, or threshold concept, 2) can provide accurate information even when majority of the vehicles are malicious or compromised.
\section{Preliminaries, Threat Models and Assumptions}
\subsection{Preliminaries}
\begin{itemize}
\item Edge Server: A trusted entity associated with a small region, such as a down-town in a city, which analyzes traffic events like accident or congestion. The edge server is responsible for accurate traffic monitoring based on the proposed heuristic, and it filters malicious data and vehicles. The use of edge server scales the proposed model for large VANETs using the distributed cloud concept.
\item Centralized Server: A trusted entity that is responsible for analyzing the traffic scenario of a city, county or a state. Since traffic congestion in a region has a cascading effect on other surrounding regions, the decisions from different edge servers deployed in small regions are analyzed by the centralized server to generate an overview of a traffic scenario for a large region.
\item Decision Similarity Graph (DSG): The Decision Similarity Graph (DSG), represented in Figure 1, is utilized by the edge server to filter malicious vehicles within the ROI, which is explained in section IV(c). DSG is an ordered pair of the form:

\hspace*{20mm}\normalfont$DSG = <V,E>$ \\ where V=$<$V\textsubscript{1}, V\textsubscript{2}, V\textsubscript{3},..., V\textsubscript{n}$>$ represent the \emph{n} vehicles from which the responses are received and E=$<$E\textsubscript{1}, E\textsubscript{2}, E\textsubscript{3},..., E\textsubscript{n}$>$ are undirected edges that represent the neighborhood between any two vehicles.
\begin{figure}[h]
      \centering
           \includegraphics[width=74mm,height=70mm,keepaspectratio]{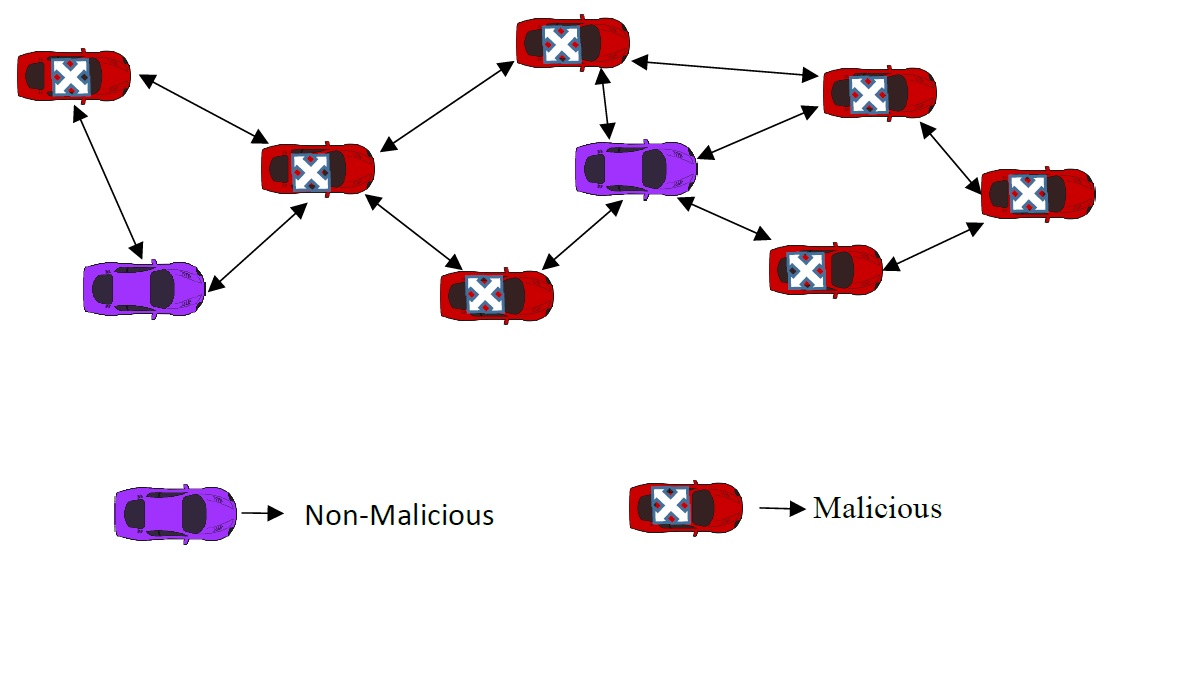}
      \caption{Decision Similarity Graph}
      \label{figurelabel1}
   \end{figure}
\item AES encryption algorithm \cite{daemen1998block}: Every vehicle, \emph{V\textsubscript{id}}, at the ROI uses the AES 128 bit symmetric encryption algorithm to generate a unique key, \emph{Key\textsubscript{i}}, which is used to encrypt the vehicular data such as ID, GPS location, and velocity before sending the data to the edge server via the RSU. Since the AES 128 bit algorithm is faster than the AES 192/256 bit key algorithms \cite{elminaam2010evaluating}, and still provide enough desired security, the proposed model reduces the latency due to encryption.

\item Schmidt-Samoa cryptosystem \cite{schmidt2006new}: Every edge server using the Schmidt-Samoa cryptosystem generates a public key, \emph{G\textsubscript{public}}, and its associated private key, \emph{G\textsubscript{private}}, which is utilized by the vehicles within the ROI for a secure key, \emph{Key\textsubscript{i}}, exchange as well as for the privacy preserved \emph{V\textsubscript{id}} broadcast to its neighboring vehicles. Edge servers associated with different regions generate their own \emph{G\textsubscript{public}} and \emph{G\textsubscript{private}} keys. We prefer the \cite{schmidt2006new} over RSA \cite{rivest1978method} because of its simplistic trapdoor one-way permutation. Also, unlike Rabin \cite{rabin1979digitalized}, this algorithm does not produce any ambiguity in the decryption at the cost of the encryption speed. The Schmidt-Samoa cryptosystem is also preferred over the elliptic-curve cryptosystem \cite{koblitz1987elliptic} because of the implementation issues faced by the latter, and also the lack of research on the latter cryptosystem \cite{magons2016applications}.

\item Elgamal Digital Signature Scheme (EGDSS) \cite{elgamal1985public}: Every vehicle, \emph{V\textsubscript{id}}, before entering into a VANET communication, is registered with the trusted centralized server using a private key, \emph{veh\_private\textsubscript{i}}, and its associated public key, \emph{veh\_public\textsubscript{i}}, which are generated using the EGDSS algorithm. The \emph{veh\_private\textsubscript{i}} is loaded onto the on-board-unit(OBU) of \emph{V\textsubscript{id}} which is used to authenticate its identity to the edge server. Instead of \emph{EGDSS}, the digital signature algorithm (DSA) \cite{kravitz1993digital} could also be used without affecting the model's effectiveness. 
\end{itemize}
\subsection{Threat Model} 
 Malicious vehicles (1) can manipulate any recorded event (including velocity and recorded GPS location) to disrupt the decision making process, (2) may try to impersonate some other vehicle by reporting some other \emph{V\textsubscript{id}}, and can also send manipulated traffic-related information under different registered \emph{V\textsubscript{id}}s, (3) may intercept a data packet of any non-malicious vehicle, modify the information, and send it to the RSU for decision making, (4) may not send its vehicular data to the RSU to degrade the traffic monitoring system.
\subsection{Assumptions}
\thispagestyle{empty}
We assume that the non-malicious vehicles always send requested information to the edge server via the trusted RSU. Every challenge and response packet is sent and received by the RSU as well as by the vehicles. If a packet originating from a vehicle is dropped due to a network problem or channel congestion, it is neglected by the proposed model. However, for the effectiveness of the proposed model, the edge server must obtain at least one non-malicious response or one conflicting neighbor of any malicious vehicle to validate the traffic-related information from the ROI if it exists. The vehicles are considered to have a unique ID (allocated by the department of motor vehicles) of uniform length known by the edge servers. The recorded event is either congested or non-congested, defined based on the velocity of the vehicles within the ROI in the proposed model (section IV c). The proposed model is not appropriate for real-time decision making such as turning the steering wheel or increasing the acceleration. It is suitable in scenarios where the requesting vehicles want to enter the ROI within approximately \emph{5-10 minutes} from the time of the request.
 %\end{itemize}
\section{Proposed Model}
Figure 3 represents an overview of the proposed model shown in Figure 2. It uses a privacy-preserving heuristic that leverages the GPS location and velocity of the reporting vehicle as well as encrypted neighboring \emph{V\textsubscript{id}} of the reporting vehicle, i.e., vehicles that are within its transmission range %(usually around 600m \cite{hsu2017transmission}), 
to validate traffic-related information at the ROI under the presence of at least one non-malicious vehicle. A vehicle requesting traffic-related information from an ROI sends a request to the centralized server, which is relayed to the associated edge server via a wireless communication. From one edge server, the requesting vehicle can request the traffic condition at another ROI under a different edge server. The centralized server is directly associated with all edge servers.
\begin{figure}[h]
      \centering
           \includegraphics[width=57 mm]{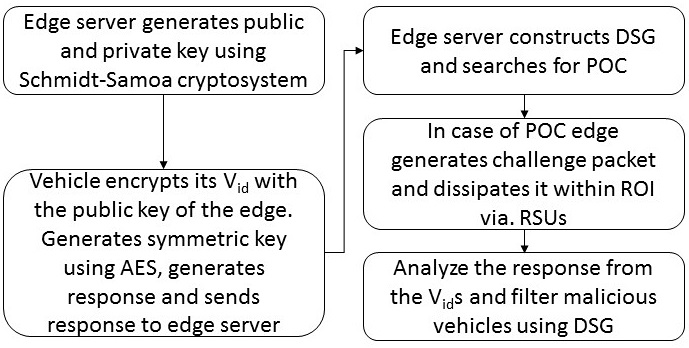}
      \caption{Overview of the proposed model}
      \label{figurelabel1}
   \end{figure}\begin{figure}[]
      \centering
           \includegraphics[width=74mm,height=70mm,keepaspectratio]{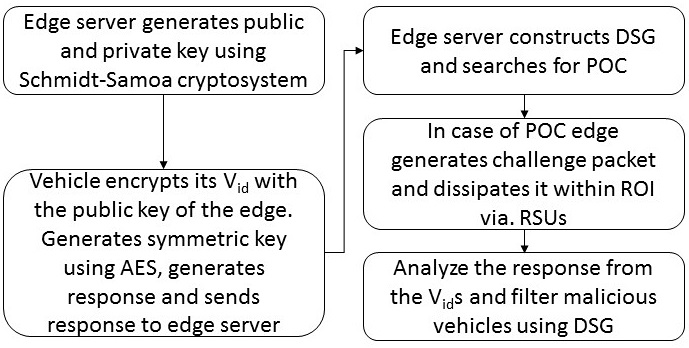}
      \caption{Overview of the proposed model}
      \label{figurelabel1}
   \end{figure}
\subsection{Key Generation and Request Dissipation}
The edge server associated with a ROI generates \emph{G\textsubscript{public}} and its associated \emph{G\textsubscript{private}} using equations 1 and 2, respectively, as defined by the Schmidt-Samoa cryptosystem.

$G\textsubscript{public} = p^2 * q $ \hspace*{57mm} (1) 

$G\textsubscript{private} = G\textsubscript{public}^{-1} \mod (lcm(p-1,q-1))$ \hspace*{12mm} (2) 

where \emph{p} and \emph{q} are 2 large prime numbers chosen by the edge server.
Furthermore, the edge server broadcasts \emph{G\textsubscript{public}} within the ROI via the RSUs.
\subsection{Encrypted ID Broadcast and Response Acquisition}
In the proposed heuristic, every vehicle needs to know the encrypted \emph{V\textsubscript{id}}s of its neighboring vehicles only when generating the \emph{data\_packet}. This requirement has been justified later in this section. To preserve the privacy in the proposed model, after receiving the \emph{G\textsubscript{public}}, the vehicle encrypts its \emph{V\textsubscript{id}} using \emph{G\textsubscript{public}} and generates \emph{enc\_id}, which is defined as the encrypted \emph{V\textsubscript{id}} as discussed in Algorithm 1 (lines 6-9), and broadcasts it to its nearby vehicles.
\begin{algorithm}[]
\caption{Generate \emph{enc\_id} and $\tau$ for a vehicle}
\begin{algorithmic}[1]
\State $key[0-9,A-Z,a-z] = [00-09,10-35,46-71]$ 
\State \emph{Key\textsubscript{i}} $\gets$ Key of a vehicle generated using AES 128
\State $Key\_m$ $\gets$ Stores integer of \emph{Key\textsubscript{i}} using key[] of step 1
\State count $\gets$ length\_of(V\textsubscript{id}) - 1
\State Key\_length $\gets$ length\_of(\emph{Key\textsubscript{i}}) - 1
\For{i in range of $0-length\_of(V\textsubscript{id})-1$}
\State $msg = msg+(100\textsuperscript{count}*$ascii\_of(V\textsubscript{id}.charAt(i)) 
\State count- -
\EndFor
\State \emph{enc\_id} $\gets$ \emph{msg}\textsuperscript{G\textsubscript{public}} mod (G\textsubscript{public})
\For{i in range of $0-length\_of(Key\textsubscript{i})$}
\State $Keym = Keym+(100\textsuperscript{Key\_length}*$key[Key\textsubscript{i}.charAt(i)] 
\State Key\_length- -
\EndFor
\State $\tau$ $\gets$ \emph{Keym}\textsuperscript{G\textsuperscript{public}} mod (G\textsubscript{public})
\State \Return \emph{enc\_id}, $\tau$
\end{algorithmic}
\end{algorithm} 
A vehicle having received \emph{threshold} number of \emph{enc\_id}s from neighboring vehicles generates \emph{data\_packet} that consists of the vehicle's information. The significance of \emph{threshold} in the proposed model is to model the size of the \emph{data\_packet} in such a way that it consumes lesser bandwidth to send the vehicular information by the RSU.

$data\_packet =<V\textsubscript{id},ds(V\textsubscript{id}),event,vel\textsubscript{id}, \\
\hspace*{30mm}GPS\textsubscript{id},enc\_ids,trajectory>$ 

where \emph{ds(V\textsubscript{id})} is defined as \emph{V\textsubscript{id}} digitally signed with \emph{veh\_private\textsubscript{i}} for authenticating itself, \emph{enc\_ids} refers to the encrypted ids of the neighboring \emph{threshold} vehicles, \emph{vel\textsubscript{id}} and \emph{GPS\textsubscript{id}} are the velocity and GPS location of the vehicle respectively at the time of generating the \emph{data\_packet}, while the \emph{event} field indicates the event recorded by the \emph{V\textsubscript{id}}. We do not preexamine the recorded GPS location of the vehicles and the proposed model does not deal with the precision of the GPS location of a \emph{V\textsubscript{id}}, as the GPS location recorded is utilized to filter malicious vehicles in the heuristic described later. The \emph{trajectory} field describes the trajectory of the vehicle to its destination that is leveraged in section D.%The precision of the GPS is insignificant in the proposed model. 
Thereafter, every \emph{V\textsubscript{id}} at the ROI generates its \emph{Key\textsubscript{i}} and generates \emph{encrypted\_data\_packet} consisting of every information that needs to be sent to the edge server in the encrypted form. 

$encrypted\_data\_packet = < \tau, data\_packet`> $

where $\tau$ is obtained by encrypting \emph{Key\textsubscript{i}} of a \emph{V\textsubscript{id}} with \emph{G\textsubscript{public}} (Algorithm 1 lines 10-13) and \emph{data\_packet`} is obtained by encrypting \emph{data\_packet} with \emph{Key\textsubscript{i}}. The purpose of this step is to preserve the privacy and integrity of \emph{Key\textsubscript{i}} and the \emph{data\_packet}. This step also facilitates the secure key exchange algorithm. Every \emph{V\textsubscript{id}} broadcasts the \emph{encrypted\_data\_packet} received by the nearby RSU. The RSU waits for $\sigma$ seconds or \emph{threshold} number of \emph{encrypted\_data\_packet}s before sending them to the edge server. The waiting time, $\sigma$, ensures that the RSU constrains the time of the proposed model in case the \emph{threshold} packets take more time, especially in situations where the traffic flow is low. On the other hand, if the traffic flow is high, within $\sigma$ time, the RSU can receive a large number of \emph{encrypted\_data\_packet}s which can significantly increase the length of the aggregated packet under which it appends \emph{threshold} \emph{encrypted\_data\_packet}s obtained before $\sigma$ seconds. Thus, using this trade-off, the RSU generates the \emph{aggregate\_packet} and sends it to the edge server. 

$aggregate\_packet = <rsu\_id, location,\\   \hspace*{35 mm}encrypted\_data\_packets>$

where \emph{rsu\_id} and \emph{location} are respectively the unique id and location of the RSU sending the packets to the edge. %server. %The use of this information for the decision making process by the edge server is explained in the subsequent subsections in this paper.  
\subsection{Point of Conflict Detection}
\thispagestyle{empty}
The edge server, on receiving the \emph{aggregate\_packet}s from the RSUs, extracts the \emph{rsu\_id} and the location of the RSU. Thereafter, it extracts the \emph{encrypted\_data\_packet} from the \emph{aggregate\_packet}s received. Furthermore, from the \emph{encrypted\_data\_packet}, the symmetric key, \emph{Key\textsubscript{i}}, for every vehicle is obtained by decrypting $\tau$ with \emph{G\textsubscript{private}} using Algorithm 2 (lines 4-5). Finally, the obtained \emph{Key\textsubscript{i}} of a vehicle is further used to obtain \emph{data\_packet} from \emph{data\_packet'} (Algorithm 2 line 6). Thus, the response of a vehicle remains private from the RSU as well as from the nearby vehicles as \emph{G\textsubscript{private}} is only possessed by the edge server. From every \emph{data\_packet}, the \emph{V\textsubscript{id}} is authenticated by the edge server by comparing \emph{ds(V\textsubscript{id})} with the \emph{V\textsubscript{id}} and its associated \emph{veh\_public\textsubscript{i}}, which it receives from the centralized server. %(generated during the vehicle registration using EGDSS) 
%as in Figure 4. 
The purpose of this step is to filter malicious vehicles that are performing any masquerading attack or identity theft. At first, the GPS location of a V\textsubscript{id} is compared with the location of the RSU. If the GPS location is out of the transmission range of the RSU that records its data, this implies that the vehicle is manipulating its GPS location and it is filtered out as malicious. If not, the neighbors of a \emph{V\textsubscript{id}} are obtained by the edge server using Algorithm 2 (lines 7-9). Based on the neighbors extracted, the edge server constructs a DSG where every \emph{V\textsubscript{id}} forms a vertex of the DSG, and it has an undirected edge to its neighboring \emph{V\textsubscript{id}}s. The undirected DSG is used to obtain the Point of Conflict (POC), defined as a situation where two neighboring vehicles, say \emph{V\textsubscript{i}} and \emph{V\textsubscript{j}}, report a conflicting event like congestion and no-congestion, respectively within the same ROI (lines 15-20) and under the same RSU. The initial detection for \emph{POC} is searched within the neighbors using Algorithm 2 (lines 10-14).
If no \emph{POC} is detected after the initial detection, a \emph{POC} can still exist among the vehicles within the same RSU (Algorithm 2 (lines 15-18)) if a malicious \emph{V\textsubscript{id}} intentionally chooses only malicious neighboring \emph{V\textsubscript{id}}s as analyzed in \textbf{Property 1}.

We consider Veh\_list\textsubscript{m} $= \{V\textsubscript{m}\textsuperscript{1},V\textsubscript{m}\textsuperscript{2},V\textsubscript{m}\textsuperscript{3},...\}$ to be a set of malicious V\textsubscript{id}, denoted by V\textsubscript{m}\textsuperscript{i}, whereas Veh\_list\textsubscript{nm} $= \{V\textsubscript{nm}\textsuperscript{1},V\textsubscript{nm}\textsuperscript{2},V\textsubscript{nm}\textsuperscript{3},...\}$ is a set of non-malicious V\textsubscript{id}, denoted by V\textsubscript{nm}\textsuperscript{i}. The cardinality of a set S is denoted by $|S|$. $\Gamma(V\textsubscript{m}\textsuperscript{i})$ represents a set of V\textsubscript{m}\textsuperscript{j}s present in the neighbor list of a V\textsubscript{m}\textsuperscript{i} and  Veh\_list\textsubscript{m} $\cap$ Veh\_list\textsubscript{nm} $= \emptyset$, i.e, a V\textsubscript{id} cannot be malicious and non-malicious at the same time.
\begin{namedtheorem}
Given a V\textsubscript{m}\textsuperscript{i}, it can have only V\textsubscript{m}\textsuperscript{j}s in its neighbor list if $threshold < |Veh\_list\textsubscript{m}|$.
\end{namedtheorem}
\begin{proof}
For $threshold < |Veh\_list\textsubscript{m}|$, consider the value of $threshold$ to be $|Veh\_list\textsubscript{m}|$-1 , i.e, the maximum allowable $threshold$ under the constraint.

For a given V\textsubscript{m}\textsuperscript{i},
V\textsubscript{m}\textsuperscript{i} $\cup$ $\Gamma(V\textsubscript{m}\textsuperscript{i}) = Veh\_list\textsubscript{m}$,

when the value of $threshold = |Veh\_list\textsubscript{m}|-1$, meaning that the neighbor list of a V\textsubscript{m}\textsuperscript{i} can include all other V\textsubscript{m}\textsuperscript{j} to avoid the detection of \emph{POC}.

V\textsubscript{m}\textsuperscript{i} $\cup$ $\Gamma(V\textsubscript{m}\textsuperscript{i}) \subset Veh\_list\textsubscript{m}$,

when the value of $threshold < |Veh\_list\textsubscript{m}|-1$, meaning that the neighbor list of a V\textsubscript{m}\textsuperscript{i} can include some V\textsubscript{m}\textsuperscript{j} to avoid the detection of \emph{POC}.

For $threshold > |Veh\_list\textsubscript{m}|$, consider the value of $threshold$ to be $|Veh\_list\textsubscript{m}|$+1 , i.e, the minimum allowable $threshold$ under the constraint.
 
$\therefore$ , $ Veh\_list\textsubscript{m} \subset V\textsubscript{m}\textsuperscript{i} \cup \Gamma(V\textsubscript{m}\textsuperscript{i})$ 

This means a V\textsubscript{m}\textsuperscript{i} will have a V\textsubscript{nm}\textsuperscript{i} in its neighbor list, under which the condition given below holds.

$V\textsubscript{m}\textsuperscript{i} \cup \Gamma(V\textsubscript{m}\textsuperscript{i}) \subset Veh\_list\textsubscript{m} \cup Veh\_list\textsubscript{nm} $
\end{proof}
\begin{algorithm}[]
\caption{Obtaining data\_packet and Detecting POC}
\begin{algorithmic}[1]
\State $key[0-9,A-Z,a-z] = [00-09,10-35,46-71]$ 
\State $dec\_id$ $\gets$ decrypted \emph{enc\_id}, POC\_detected $\gets$ false
\State neighbor\_of\_V\textsubscript{i} $\gets$ neighboring vehicle of a V\textsubscript{id}
\For {every $\tau$ received}
\State \emph{Key1\textsubscript{i}} $\gets$ key[$\tau$\textsuperscript{G\textsubscript{private}} mod (p*q)]
\EndFor
\State data\_packet $\gets$ decrypt $data\_packet`$ with Key1\textsubscript{i}
\For {every \emph{enc\_id} in \emph{data\_packet}}
\State $dec\_id \gets$ enc\_id$\textsuperscript{G\textsubscript{private}} mod (p*q)$
%\State $count \gets $ length\_of(V\textsubscript{id}) - 1
%\State neighbor\_of\_V\textsubscript{i} $\gets$ ascii\_string\_of{\emph{enc\_id}\textsuperscript{G\textsubscript{private}} mod (p*q)}
%\While {count- - $<0$}
\State neighbor\_of\_V\textsubscript{i} $\gets$ ascii\_characters\_of(\emph{dec\_id})
%\EndWhile
\EndFor
%\State POC\_detected $\gets$ false
\For{every data\_packet obtained from V\textsubscript{i}}
\For{every V\textsubscript{j} in neighbor\_of\_V\textsubscript{i}}
\If {V\textsubscript{i}.event $\neq$ V\textsubscript{j}.event}
\State POC\_detected $=$ true, CV\textsubscript{1} $\gets$ V\textsubscript{i}, CV\textsubscript{2} $\gets$ V\textsubscript{j}
%\State CV\textsubscript{1} $\gets$ V\textsubscript{i}
%\State CV\textsubscript{2} $\gets$ V\textsubscript{j}
\State \textbf{go to} Line 19
\EndIf
\EndFor
\EndFor
\For{any 2 vehicles, V\textsubscript{k} and V\textsubscript{m} under same rsu\textsubscript{id}}
\If {V\textsubscript{k}.event $\neq$ V\textsubscript{m}.event}
\State POC\_detected $=$ true, CV\textsubscript{1} $\gets$ V\textsubscript{k}, CV\textsubscript{2} $\gets$ V\textsubscript{m}
%\State CV\textsubscript{1} $\gets$ V\textsubscript{k}
%\State CV\textsubscript{2} $\gets$ V\textsubscript{m}
\State \textbf{go to} Line 19
\EndIf
\EndFor
\State \Return POC\_detected, CV\textsubscript{1}, CV\textsubscript{2}
\end{algorithmic}
\end{algorithm}

Thereafter, if no POC is detected at all, it considers the similar event recorded by all the vehicles to be the event of the ROI. However, if a POC is detected, the edge performs initial scrutiny based on the information obtained from the vehicles in conflict as described below. During the initial scrutiny, the approximate velocity of a vehicle in a congested road is considered \emph{vel\textsubscript{congested}}, while the velocity in a non-congested road is considered \emph{vel\textsubscript{n-congested}}, with an allowable difference of $\epsilon$ mph to accommodate any minor variations. %in velocity
 
\begin{enumerate}
\item If \emph{event} recorded by a vehicle, say \emph{V\textsubscript{1}}, is \emph{congested}, and its corresponding \emph{vel\textsubscript{1}} is greater than \emph{vel\textsubscript{congested}}+$\epsilon$, then \emph{V\textsubscript{1}} is considered malicious, and the \emph{V\textsubscript{id}} in conflict with \emph{V\textsubscript{1}} is considered non-malicious.
\item If \emph{event} recorded by a vehicle, say \emph{V\textsubscript{1}}, is \emph{non-congested}, and its corresponding \emph{vel\textsubscript{1}} is less than \emph{vel\textsubscript{n-congested}}$-\epsilon$, then \emph{V\textsubscript{1}} is considered malicious, while the \emph{V\textsubscript{id}} in conflict with \emph{V\textsubscript{1}} is considered non-malicious. 
\end{enumerate}
Subsequently, the malicious vehicles are filtered from the network using Algorithm 4 (lines 10-12), and the decision is made based on the non-malicious vehicles.
However, if no decision is made after initial scrutiny, the server generates a \emph{challenge\_pkt} obtained using Algorithm 3 (line 17), consisting of \emph{V\textsubscript{id}}s in conflict, and the RSUs under which the conflicting V\textsubscript{id}s are expected to appear after \emph{time\textsubscript{id}} is calculated based on the \emph{vel\textsubscript{id}}s, the \emph{trajectory} of the vehicle and the GPS\textsubscript{id}s of the \emph{V\textsubscript{id}}s. The purpose of the \emph{challenge\_pkt} is to authenticate the \emph{vel\textsubscript{id}} and \emph{GPS\textsubscript{id}} recorded by the vehicle and to allow the \emph{V\textsubscript{id}}s to prove its event recorded as it is assumed to travel with almost the same \emph{vel\textsubscript{id}}. 

$challenge\_pkt = < CV\textsubscript{1},expected\_rsu\textsubscript{1},
time\textsubscript{1}, \\
\hspace*{30mm}CV\textsubscript{2},expected\_rsu\textsubscript{2},time\textsubscript{2}> $

\begin{algorithm}[]
\caption{Generating the $challenge\_packet$ }
\begin{algorithmic}[1]
\State Initialize time $t$, RSUList $\gets$ list of every rsu\_id
%\State RSUList $\gets$ list of every rsu\_id
\For {every CV\textsubscript{i}}
\State calculated\_distance $\gets$ $t*vel\textsubscript{id}$
\State expected\_location\textsubscript{i} $\gets$ calculated\_distance + GPS\textsubscript{id}
\For{every $rsu\textsubscript{i} \in RSUList$ in $trajectory$}
\If {expected\_location\textsubscript{i} is within rsu\textsubscript{i}.location }
\If {CV\textsubscript{i}.event = "congested"}
\State $time\textsubscript{i} = t$, $ expected\_rsu\textsubscript{i} = rsu\textsubscript{i}$
%\State $ expected\_rsu\textsubscript{i} = rsu\textsubscript{i}$
\State \textbf{go to} Line 17
\Else
\State $ expected\_rsu\textsubscript{i} = rsu\textsubscript{i}$
\For{every rsu\textsubscript{j} beyond rsu\textsubscript{i}}
\State $time\textsubscript{i}=t$, $expected\_rsu\textsubscript{i} =expected\_rsu\textsubscript{i}+rsu\textsubscript{j}$
%\State $ expected\_rsu\textsubscript{i} =expected\_rsu\textsubscript{i}+rsu\textsubscript{j}$
\EndFor
\State \textbf{go to} Line 17
\EndIf
\Else
\State $t++$
\EndIf
\EndFor
\EndFor
\State \emph{challenge\_packet} = CV\textsubscript{i}+time\textsubscript{i}+expected\_rsu\textsubscript{i}/s
\State \Return \emph{challenge\_packet}
\end{algorithmic}
\end{algorithm} 
Based on the contents of the \emph{challenge\_pkt}s, the \emph{expected\_rsu\textsubscript{1}}  should obtain a response from \emph{CV\textsubscript{1}}, i.e. the ID of one of the vehicles in conflict,  after time,    \emph{time\textsubscript{1}}, while the \emph{expected\_rsu\textsubscript{2}} should obtain a response from \emph{CV\textsubscript{2}}, i.e. the ID of the other vehicle in conflict, after time, \emph{time\textsubscript{2}}.
To handle the case of overspeeding by a  \emph{CV\textsubscript{i}} recording "non-congested", every RSU along the direction of \emph{CV\textsubscript{i}}, obtained from its velocity, dissipates the \emph{challenge\_pkt} (Algorithm 3 lines 12-14). This ensures that even if a \emph{CV\textsubscript{i}} passed by \emph{expected\_rsu\textsubscript{i}} before \emph{time\textsubscript{i}}, it still gets the challenge packet, as over speeding is only possible in a \emph{non-congested road}.

\subsection{Challenge Packet Dissemination and Response}
\thispagestyle{empty}
Upon receiving a \emph{challenge\_pkt} from the edge server, an RSU generates a \emph{crypto\_challenge} packet and broadcasts it after \emph{time\textsubscript{i}}. The \emph{crypto\_challenge} is generated based on the \emph{CV\textsubscript{i}} assigned to it, and its purpose is to verify the presence of a vehicle within a specific region after \emph{time\textsubscript{i}} seconds. This is leveraged in the proposed heuristic to filter the malicious vehicles within the network. Every vehicle at the ROI sends a \emph{crypto\_response}, defined as the unique response sent by \emph{V\textsubscript{id}} in response to the \emph{crypto\_challenge} packet. 
The \emph{crypto\_challenge} packet is obtained by using bitwise manipulation (\emph{left shift} operation) over the \emph{XOR} cipher technique. The \emph{XOR} cipher technique is computationally inexpensive and easy to implement. Furthermore, since every RSU dissipates the \emph{crypto\_challenge} packet exactly once in the proposed model, it is less susceptible to frequency analysis attacks and also man-in-the-middle attacks. The bitwise manipulation is to enhance the security of the \emph{crypto\_challenge} after the XOR operation. 

$\hspace*{10mm}crypto\_challenge = CV\textsubscript{i} \oplus (testing\_word << left\textsubscript{num}) $
$left\textsubscript{num}$ $\in$ [1, length of $testing\_word -1$]

where \emph{testing\_word} is any arbitrary word chosen by a RSU having the same length as its assigned \emph{CV\textsubscript{i}} and $left\textsubscript{num}$ refers to the number of left shift operations performed, which is chosen arbitrarily by the RSU.  

Thereafter, upon receiving the \emph{crypto\_challenge} packet from the nearby RSU, every vehicle generates the \emph{crypto\_response} packet and broadcasts it. The purpose of the \emph{crypto\_response} packet is to validate the presence of \emph{CV\textsubscript{i}} at a specific location.

$\hspace*{10mm}crypto\_response =  V\textsubscript{id} \oplus crypto\_challenge$

%\begin{algorithm}[]
%\caption{Analyzing \emph{crypto\_response} packets}
%\begin{algorithmic}[1]
%\State Initialise response $\gets$ not received
%\State counter $\gets$ 0
%\For{{every crypto\_response received  by expected\_rsu\textsubscript{i}}}
%\If {testing\_word == (crypto\_response $>>$ left\textsubscript{num})}
%\State response $=$ received
%\State break
%\EndIf
%\EndFor
%\State vehicle\_search = rsu\_id + "," + CV\textsubscript{i}+ "," + response
%\State \Return vehicle\_search 
%\end{algorithmic}
%\end{algorithm}

The \emph{expected\_rsu\textsubscript{i}}s waits for additional $\sigma$ seconds to receive the \emph{crypto\_response} packets from the \emph{V\textsubscript{id}}s. This is done to make reparation for a minor change in \emph{vel\textsubscript{CV\textsubscript{i}}} that may occur due to any trivial circumstance that does not affect the event at the ROI. However, it is assumed that \emph{vel\textsubscript{CV\textsubscript{i}}} changes by a factor of atmost $\epsilon$ that still adheres to the decision recorded by \emph{CV\textsubscript{i}}. Thereafter, the RSU compares the \emph{crypto\_response} received from every V\textsubscript{id}s with the \emph{testing\_word}. The \emph{testing\_word} only matches with a specific \emph{CV\textsubscript{i}}. The associative and commutative nature of the XOR operation facilitates the effective analysis of the \emph{crypto\_response}s obtained.

Finally, the \emph{expected\_rsu\textsubscript{i}}s generates the \emph{vehicle\_search} packet, which is sent to the edge server. The \emph{vehicle\_search} packets report whether a \emph{CV\textsubscript{i}} was present within the transmission range of \emph{expected\_rsu\textsubscript{i}} within $time\textsubscript{i}+\sigma$.

$\hspace*{4mm}vehicle\_search = <rsu\textsubscript{id},CV\textsubscript{i},response>$

where \emph{response} can be \emph{received} indicating that a \emph{CV\textsubscript{i}} is present within the range of \emph{expected\_rsu\textsubscript{i}} within $time\textsubscript{i}+\sigma$, or \emph{not received} which indicates that the vehicle was absent.
\subsection{Decision Making by Edge Server}
The edge server, on receiving the \emph{vehicle\_search} packets from \emph{expected\_rsu\textsubscript{i}}s, makes a decision about traffic conditions and filters malicious vehicles based on the heuristic depicted in Figure 4. According to the proposed heuristic:
\thispagestyle{empty}
\begin{itemize}
\item If \emph{crypto\_response} has been received from one \emph{CV\textsubscript{i}} and is not received from the conflicting \emph{CV\textsubscript{i}}, then the \emph{CV\textsubscript{i}} from which the \emph{crypto\_response} has been received is considered non-malicious and the conflicting \emph{CV\textsubscript{i}} is considered malicious. Consequently, all the malicious vehicles with similar events recorded as the malicious \emph{CV\textsubscript{i}} are filtered using Algorithm 4 (lines 10-12). The decision is made based on the decision of the non-malicious \emph{CV\textsubscript{i}}.
\item If \emph{crypto\_response} has been received by both \emph{CV\textsubscript{i}}s, then the edge server assumes that the \emph{CV\textsubscript{i}} providing \emph{crypto\_response} with low \emph{vel\textsubscript{i}} intentionally reduced its velocity to prove itself non-malicious. Under such a scenario, the decision made is "non- congested". Thereafter, every malicious vehicle is filtered using Algorithm 4 (lines 10-12).
\end{itemize}%\vspace*{-1em}
\begin{algorithm}[]
\caption{Filtering using DSG}
\begin{algorithmic}[1]
\State CV\textsubscript{i} $\gets$ malicious vehicle detected, stack.push(CV\textsubscript{i})
\State mal\_list $\gets$ malicious V\textsubscript{id} list, mal\_list.append(CV\textsubscript{i})
%\State stack.push(CV\textsubscript{i})
%\State mal\_list.append(CV\textsubscript{i})
\State nonmal\_list $\gets$ non malicious V\textsubscript{id} list
\State filter[V\textsubscript{id}s] = false, filter[CV\textsubscript{i}] = true
%\State filtered[CV\textsubscript{i}] = true
\While{stack not empty}
\State CV\textsubscript{i}=stack.pop()
\For{every V\textsubscript{id} in DSG}
\If {hasEdge( V\textsubscript{id} , CV\textsubscript{i}) and filter[V\textsubscript{id}]==false}
\State stack.push(V\textsubscript{id})
\If {CV\textsubscript{i}.event==V\textsubscript{id}.event}
\If {CV\textsubscript{i}== malicious}
%\State V\textsubscript{id} $\gets$ malicious
\State mal\_list.append(V\textsubscript{id})
\Else
%\State V\textsubscript{id} $\gets$ non malicious
\State nonmal\_list.append(V\textsubscript{id})
\EndIf
\Else
%\State V\textsubscript{id} $\gets$ non malicious
\State nonmal\_list.append(V\textsubscript{id})
\EndIf
\EndIf
\EndFor
\EndWhile
\State \Return mal\_list, nonmal\_list
\end{algorithmic}
\end{algorithm}
%\vspace*{-1.0 em}
\begin{comment}
\end{comment}
\begin{figure}[]
      \centering 
    \includegraphics[width=\linewidth]{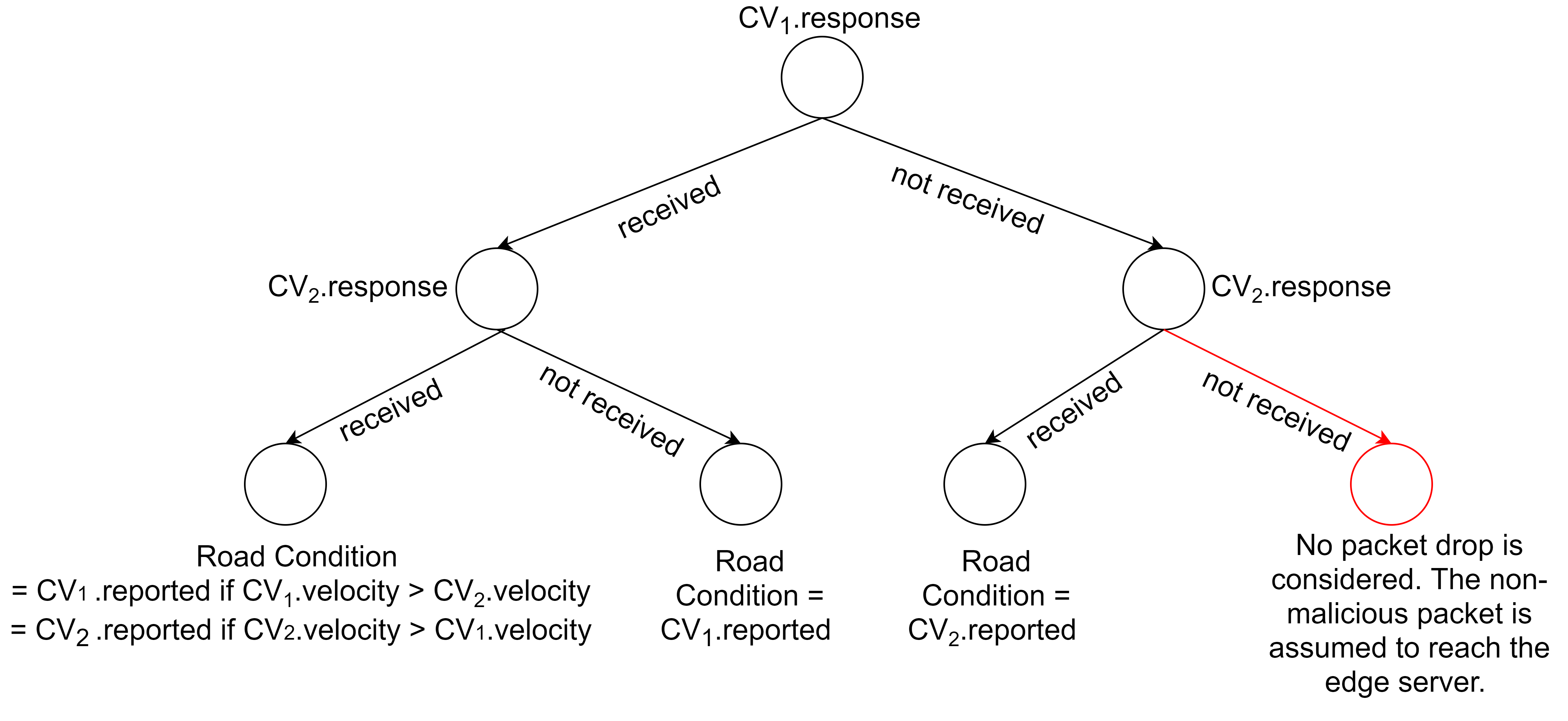}
        \caption{Decision Tree for analysis}
        \label{figurelabel}
      \end{figure}
\subsection{Response from Edge to Centralized Server}
The traffic-related information from the ROI that has been identified by the edge servers is sent to the centralized server. The centralized server sends traffic-related information to the requesting vehicle in case of an ad hoc request, or the information is stored for traffic monitoring. The centralized server observes the traffic using the majority selection method and decides that the traffic scenario of multiple regions is the one reported by the majority of the edge servers. This is because the traffic condition in one region may have a cascading effect on the other regions. The centralized server in the proposed model is free from any bottleneck issues, which may have resulted on account of enormous requests about different ROIs. In the proposed model this is handled by different edge servers, and it also allows graceful degradation as the entire system will not falter if one edge server is faulty.
\section{Experimental Results and Analysis}
\thispagestyle{empty}
%\subsection{Experimental Results}
Experiments were conducted under both congested and  non-congested road conditions. The road network was simulated using SUMO \cite{behrisch2011sumo}, while the communication network was simulated using Omnet++ and VENTOS simulators. The centralized server (cloud) and the edge server were simulated in separate workstations. In our simulation, we assumed roads that have a speed limit between \emph{70} mph and \emph{40} mph. In other words, the roads where vehicles travel below \emph{35} mph are assumed to be \emph{congested} whereas the roads where vehicles travel within the speed limit are \emph{non-congested}.
Based on the experiments performed under the parameters in Table I, it was observed that with the value of $\sigma$ that were less than 10 seconds, the proposed model obtained the packets as desired. The responses of any of the \emph{CV\textsubscript{i}} did not reach the nearby \emph{RSU} within time with any values of $\sigma$ lesser than 10 seconds. This is because on many occasions the channel remained busy transferring packets and because of this, the \emph{crypto\_response} had to wait in the pipeline as the simulator prevents any packet collision in the channel. 
The value of $\epsilon$ in the proposed model can be decided based on where it is deployed and can be adjusted without any change in the performance of the model. However, for the experimentation, we considered the value to be \emph{5} mph.
%\vspace*{-0.7 em}
\begin{table}[h]
\begin{center}
\caption {Simulation Environment and Parameters}
\label{output_1}
\resizebox{\linewidth}{!}{%
\begin{tabular} {| c || c | }
\hline
Consecutive RSU distance & 1000 $\sim$ 2500 meters \\
\hline
RSU interference distance & 510.5 meters(default simulator value)\\
\hline
Vehicles at $ROI$ & 100 \\
\hline 
Value of $\sigma$ & 10\\
\hline 
Value of \emph{threshold} & 3 vehicle data \\
\hline
Transmission Power & 20mW (default simulator value) \\
\hline
Vehicle transmission range & 510.5 meters (default simulator value) \\
\hline
Communication Protocol & IEEE 802.11p (default simulator value) \\
\hline
Communication channel & Dedicated Short Range Communication (default simulator value) \\
\hline
\end{tabular}}
\end{center}
\end{table}
The proposed model has been compared against various other models using \emph{Detection\_Accuracy} as represented in Figure 6. \emph{Detection\_Accuracy} is a metric that is used to detect accurate road conditions under a varied percentage of malicious vehicles. 

$Detection\_Accuracy = \frac{road\_condition}{\%\_of\_malicious\_vehicles}$ 

where \emph{Detection\_Accuracy} $\in$ \{1,0,0.5\}

A value of \emph{1} indicates that the accurate road condition and the malicious vehicles are detected while \emph{0} indicates neither the road condition nor the malicious vehicles could be detected. A value of \emph{0.5} indicates that the decision making is conditional, and it is either dependent on the prior reputation of vehicles (in the case of a reputation based system), the distribution of malicious vehicles (in the case of a peer authentication system) or has an equal percentage of malicious as well as non-malicious vehicles (in the case of a majority voting approach).

From Figure 6, it can be seen that the majority voting model and the peer authentication model have higher \emph{Detection\_Accuracy} when the majority of the vehicles within the ROI are non-malicious (greater than 50\%). Under such cases, the decision of the non-malicious vehicles that are in the majority dominates the decision of the malicious vehicles. However, it becomes conditional when the number of malicious and non-malicious vehicles within the ROI are equal, as the non-malicious vehicles do not form a majority, and hence no proper decision could be made. Thereafter, as the number of malicious vehicles increases beyond \emph{50\%}, it is seen that the \emph{Detection\_Accuracy} decreases as the malicious vehicles form the majority under such scenarios. This influences the decision making process within the ROI. The \emph{Detection\_Accuracy} of the \emph{Reputation\_Based} model is always 0.5. This is because the decision is highly based on the reputation of the vehicles within the ROI. It is possible to have less malicious vehicles with a higher prior reputation (for instance, 5 malicious vehicles out of 100 vehicles with high rating) to dominate the majority of non-malicious vehicles with no prior reputation (for instance, 95 non-malicious vehicles out of 100 vehicles with no reputation). Thus, under such a system, the decision making model remains conditional.
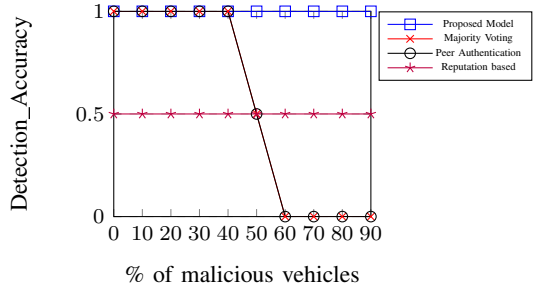
\begin{figure}[h]
    \centering
    \begin{tikzpicture}
\begin{axis}[
    footnotesize,
      title={},
      legend style={nodes={scale=0.5, transform shape}}, 
       % legend image post style={mark=*},
    xlabel={\% of malicious vehicles},
    ylabel={Detection\_Accuracy},
    xmin=0, xmax=90,
    ymin=0, ymax=1,
    xtick={0,10,20,30,40,50,60,70,80,90},
    ytick={0,0.5,1},
    legend pos= outer north east,
    ymajorgrids=true,
    legend entries={Proposed Model,
                Majority Voting,
                Peer Authentication,
                Reputation based},
    grid style=dashed,
]
\addplot[
    color=blue,
    mark=square,
    ]
    coordinates {
    (0,1)(10,1)(20,1)(30,1)(40,1)(50,1)(60,1)(70,1)(80,1)(90,1)
    }; %\addlegendentry[text width=45pt, text depth=]{Proposed Model}
 
\addplot[
    color=red,
    mark=x,
    ]
    coordinates {
    (0,1)(10,1)(20,1)(30,1)(40,1)(50,0.5)(60,0)(70,0)(80,0)(90,0)
    }; %\addlegendentry[text width=45pt, text depth=]{Majority Voting}

\addplot[
    color=black,
    mark=o,
    ]
    coordinates {
    (0,1)(10,1)(20,1)(30,1)(40,1)(50,0.5)(60,0)(70,0)(80,0)(90,0)
    }; %\addlegendentry[text width=45pt, text depth=]{Peer Authentication}

\addplot[
    color=purple,
    mark=star,
    ]
    coordinates {
    (0,0.5)(10,0.5)(20,0.5)(30,0.5)(40,0.5)(50,0.5)(60,0.5)(70,0.5)(80,0.5)(90,0.5)
    }; %\addlegendentry[text width=45pt, text depth=]{Reputation based model}
    
\end{axis}
\end{tikzpicture}
\caption{Comparison of Detection Accuracy of various models}
    \label{fig:my_label}
\end{figure}
It is also to be noted that the number of broadcasts required in the proposed model by the vehicles is comparatively less compared to the \emph{peer authentication} model, as shown in Figure 7. Furthermore, we see that the peer authentication model has an equal number of broadcasts per vehicle compared to the \emph{Proposed Model Lower}, especially%that involves the number of broadcasts required for decision making 
during the initial scrutiny phase when the \emph{threshold} $\leq2$. However, the number of broadcasts required for decision making in the peer authentication model becomes equal to the \emph{Proposed Model Upper} if the \emph{challenge\_response} packet is generated with $threshold=3$, and it exceeds the \emph{Proposed Model Upper} when the threshold becomes $>3$. This is because every vehicle has to authenticate \emph{threshold} number of vehicles in the peer authentication model, which increases with the increase in \emph{threshold} and with the number of vehicles at the ROI. The value of \emph{threshold} represents the number of neighboring vehicles that needs to be authenticated by a vehicle as well as the number of vehicular response sent by the RSU to the edge in one time period. Since the proposed model involves no \emph{peer authentication}, every vehicle within the ROI has to broadcast twice during the initial scrutiny phase, and thrice if \emph{challenge\_response} is generated. However, the proposed model has more broadcast messages per vehicle compared to the \emph{majority voting} and the \emph{reputation-based} model. This is because the \emph{majority voting} model and the \emph{reputation-based} model involve no \emph{V2V} communication, and every vehicle has to broadcast its decision only once. Furthermore, we formulate the total broadcast required for the various models shown in Table II, where \emph{n} represents the number of vehicles within the ROI and  \emph{n\textsubscript{rsu}} represents the number of RSUs deployed within the ROI.

In the case of majority voting and reputation-based models, every vehicle within the ROI sends their response to the RSU that has a total of \emph{n} transmissions. Thereafter, the RSU sends \emph{threshold} number of vehicular responses to the edge server at one time. Hence, all the packets are sent after $\frac{n}{threshold}$ times.
Therefore, the total number of transmissions required is \emph{n+$\frac{n}{threshold}$}.

In the peer authentication model, every vehicle at the ROI authenticates \emph{threshold} number of vehicles. Therefore, the number of authentication transmissions are $n*threshold$. Thereafter, the RSU sends all the packets to the edge server after $\frac{n*threshold}{threshold}$ times, i.e., \emph{n} times. Thus, the total number of transmissions required in the peer authentication model is $(n*threshold)+n$, which is equivalent to $n*(1+threshold)$.

In our proposed model, every vehicle sends its \emph{enc\_id} to its neighbor and thereafter, it sends the \emph{encrypted\_packet} to the RSU. Therefore, for every vehicle, it involves 2 transmissions. For \emph{n} vehicles, the total transmissions are $2*n$. The transmissions required by RSUs to send \emph{n} packets to the edge server is $\frac{n}{threshold}$. Thereafter, the initial scrutiny is performed. If the decision is made after initial scrutiny using the proposed model, we obtain the lower bound, \emph{Proposed Lower}, on the transmissions, which is $(2*n)+\frac{n}{threshold}$. However, when the \emph{challenge\_response} phase is executed, the challenge packet is sent to the \emph{n\textsubscript{rsu}} RSUs by the edge server in one transmission, which is dissipated within the ROI. \emph{n\textsubscript{rsu}} RSUs broadcast the \emph{crypto\_challenge} packet and receive the \emph{crypto\_response} packets from \emph{n} vehicles. Therefore, the total transmissions are $n+n\textsubscript{rsu}+1$. Finally, \emph{n\textsubscript{rsu}} RSUs send their response to the edge server in \emph{n\textsubscript{rsu}} transmissions. Under the \emph{challenge\_response} phase, we obtain the upper bound on the transmissions of the proposed model, \emph{Proposed Upper}, which is $2*n+\frac{n}{threshold}+2*n\textsubscript{rsu}+n+1$.
Based on the formulation, we find from Figure 8 that the total energy required for the transmission by the majority voting and reputation-based models are the least compared to the proposed model and the peer authentication model. This is because they involve no \emph{V2V} communication. Therefore, with fewer number of broadcast, the transmission energy used is also less. However, our proposed model has fewer number of broadcast in both initial scrutiny (Proposed Lower) and when the challenge\_response is generated (Proposed Upper) than the peer authentication model because every vehicle has a fewer broadcast requirement, i.e., two broadcasts in \emph{Proposed Lower} and three broadcasts in \emph{Proposed Upper}. Hence, it requires less transmission energy compared to the peer authentication model, where every vehicle has to authenticate \emph{threshold} number of neighboring vehicles.
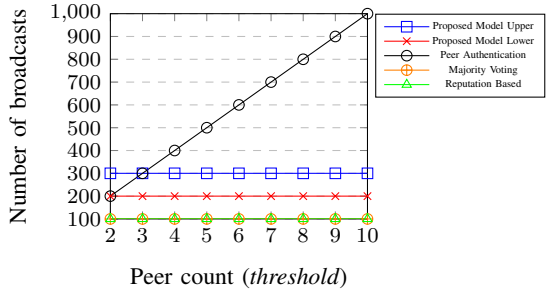
\begin{figure}[h]
    \centering
 %    \resizebox {0.45\textwidth} {!} {
    \begin{tikzpicture}
\begin{axis}[
    footnotesize,
      title={},
      legend style={nodes={scale=0.5, transform shape}}, 
       % legend image post style={mark=*},
    xlabel={Peer count (\emph{threshold})},
    ylabel={Number of broadcasts},
    xmin=2, xmax=10,
    ymin=100, ymax=1000,
    xtick={2,3,4,5,6,7,8,9,10},
    ytick={100,200,300,400,500,600,700,800,900,1000},
    legend pos= outer north east,
    ymajorgrids=true,
    legend entries={Proposed Model Upper,
                Proposed Model Lower,
                Peer Authentication,
                Majority Voting,
                Reputation Based},
    grid style=dashed,
]

%\addlegendimage{empty legend}
   % \addlegendentry{\hspace{-1cm}\textbf{Models}}
 
\addplot[
    color=blue,
    mark=square,
    ]
    coordinates {
    (2,300)(3,300)(4,300)(5,300)(6,300)(7,300)(8,300)(9,300)(10,300)
    }; %\addlegendentry[text width=45pt, text depth=]{Proposed Model}
 
\addplot[
    color=red,
    mark=x,
    ]
    coordinates {
    (2,200)(3,200)(4,200)(5,200)(6,200)(7,200)(8,200)(9,200)(10,200)
    }; %\addlegendentry[text width=45pt, text depth=]{Majority Voting}

\addplot[
    color=black,
    mark=o,
    ]
    coordinates {
    (2,200)(3,300)(4,400)(5,500)(6,600)(7,700)(8,800)(9,900)(10,1000)
    }; %\addlegendentry[text width=45pt, text depth=]{Peer Authentication}

\addplot[
    color=orange,
    mark=oplus,
    ]
    coordinates {
    (2,100)(3,100)(4,100)(5,100)(6,100)(7,100)(8,100)(9,100)(10,100)
    }; %\addlegendentry[text width=45pt, text depth=]{Proposed Model}

\addplot[
    color=green,
    mark=triangle,
    ]
    coordinates {
    (2,100)(3,100)(4,100)(5,100)(6,100)(7,100)(8,100)(9,100)(10,100)
    }; %\addlegendentry[text width=45pt, text depth=]{Proposed Model}
\end{axis}
\end{tikzpicture}
\caption{Broadcast comparison for 100 vehicles with varying threshold}
    \label{fig:my_label}
\end{figure}
%\vspace*{-1.5em}
\begin{table}[h]
\begin{center}
\caption {Total number of broadcasts}
\label{output_1}
\resizebox{\linewidth}{!}{%
\begin{tabular} {| c || c | }
\hline
\scriptsize{Majority Voting} & \scriptsize{(n+$\frac{n}{threshold}$)} \\
\hline
\scriptsize{Reputation Based} & \scriptsize{(n+$\frac{n}{threshold}$)}\\
\hline
\scriptsize{Peer Authentication} & \scriptsize{(n*(1+threshold))} \\
\hline 
\scriptsize{Proposed Lower} & \scriptsize{(2*n+$\frac{n}{threshold}$)}\\
\hline 
\scriptsize{Proposed Upper} & \scriptsize{(2*n+$\frac{n}{threshold}$+2*n\textsubscript{rsu}+n+1)}\\
\hline
\end{tabular}}
\end{center}
\end{table}
\thispagestyle{empty}
%\vspace*{-1.0em}
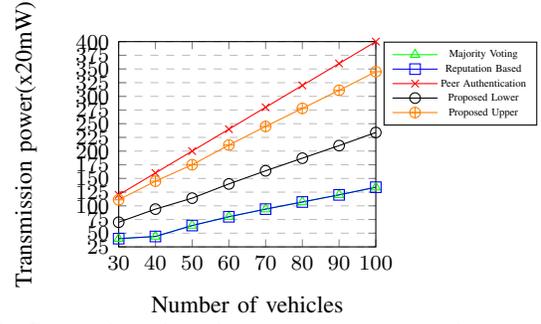
\begin{figure}[h]
    \centering
 %    \resizebox {0.40\textwidth} {!} {
    \begin{tikzpicture}
\begin{axis}[
    footnotesize,
      title={},
      legend style={nodes={scale=0.5, transform shape}}, 
       % legend image post style={mark=*},
    xlabel={Number of vehicles},
    ylabel={Transmission power(x20mW)},
    xmin=30, xmax=100,
    ymin=25, ymax=400,
    xtick={30,40,50,60,70,80,90,100},
    ytick={25,50,75,100,125,150,175,200,225,250,275,300,325,350,375,400},
    legend pos= outer north east,
    ymajorgrids=true,
    legend entries={Majority Voting,
                    Reputation Based,
                    Peer Authentication,
                    Proposed Lower,
                    Proposed Upper},
    grid style=dashed,
]

\addplot[
    color=green,
    mark=triangle,
    ]
    coordinates {
    (30,40)(40,44)(50,64)(60,80)(70,94)(80,107)(90,120)(100,134)
    }; %\addlegendentry[text width=45pt, text depth=]{Proposed Model}

\addplot[
    color=blue,
    mark=square,
    ]
    coordinates {
    (30,40)(40,44)(50,64)(60,80)(70,94)(80,107)(90,120)(100,134)
    }; %\addlegendentry[text width=45pt, text depth=]{Proposed Model}
 
\addplot[
    color=red,
    mark=x,
    ]
    coordinates {
    (30,120)(40,160)(50,200)(60,240)(70,280)(80,320)(90,360)(100,400)
    }; %\addlegendentry[text width=45pt, text depth=]{Majority Voting}

\addplot[
    color=black,
    mark=o,
    ]
    coordinates {
    (30,70)(40,94)(50,114)(60,140)(70,164)(80,187)(90,210)(100,234)
    }; %\addlegendentry[text width=45pt, text depth=]{Peer Authentication}

\addplot[
    color=orange,
    mark=oplus,
    ]
    coordinates {
    (30,111)(40,145)(50,175)(60,211)(70,245)(80,278)(90,311)(100,345)
    }; %\addlegendentry[text width=45pt, text depth=]{Proposed Model}
\end{axis}
\end{tikzpicture}
\caption{Comparison based on energy consumption per transmission}
    \label{fig:my_label}
\end{figure}
The time taken by our proposed model is highly dependant on the \emph{POC} detection. It is noted from Figure 9 (plotted based on Table III) that the detection time for our model increases with the \emph{POC} distance (the point where two vehicles conflict in event reporting as detected by the edge). This is because only when the \emph{POC} is detected, the edge server begins the various steps in our proposed model for detecting the road condition, and filtering out malicious vehicles. When the \emph{POC} is detected early, i.e., within less \emph{POC} distance, the edge proceeds with the initial scrutiny and thereafter, the \emph{challenge\_response} packet may be generated. However, with the increase in the \emph{POC} distance, the edge server has to wait longer before executing the initial scrutiny phase. It can also be seen from Table III that the time taken is less dependant on the consecutive RSU distances compared to the \emph{POC} distances during \emph{Proposed Lower}. This is because even if the RSUs are close to each other, the edge server still has to detect a \emph{POC} before the initial scrutiny phase is executed. The \emph{Proposed Upper} with a consecutive RSU distance of \emph{2500} and a \emph{POC} distance of 10 in Table III takes less time (79 seconds) compared to the \emph{Proposed Lower} with a consecutive RSU distance of \emph{1000} and a \emph{POC} distance of 20 (82.5 seconds). This is because in the latter scenario, the initial \emph{POC} takes time to be detected by the edge. Therefore, a scenario having RSUs close to each other with a higher \emph{POC} distance may take more time than a scenario with relatively distant consecutive RSUs but with less \emph{POC} distance. Also, in the initial scrutiny phase, the time taken does not depend on the RSU distances (all the \emph{Lower} values %, i.e \emph{2500 Lower}, \emph{2000 Lower}, \emph{1500 Lower}, \emph{1000 Lower} 
from Table III take the same time for detection for a given \emph{POC} distance). This is because the vehicle responses is leveraged to make the decision after they are sent to the edge server. The RSU distance is impactful only when the \emph{challenge\_response} packet is generated (all the \emph{Upper} values %, i.e \emph{2500 Upper}, \emph{2000 Upper}, \emph{1500 Upper}, \emph{1000 Upper} 
from Table III take different times for detection for a given \emph{POC} distance). Through experiments, it was observed that the proposed model performs faster when the vehicles remain under some RSU throughout their travel. This is because if the vehicles are out of the transmission range of the RSU, it has to wait to arrive near the next RSU before sending their corresponding packets, which increases the latency. In the experiments, every vehicle remained under the transmission range of the RSU throughout their travel when they were placed \emph{1000} metres apart, and it was observed that \emph{1000 Upper} as shown in Table III has the fastest performance when compared to \emph{2500 Upper}, \emph{2000 Upper}, and \emph{1500 Upper}, i.e., when the consecutive RSU distances were \emph{2500 metres}, \emph{2000 metres}, and \emph{1500 metres} apart, respectively. In Figure 9, \emph{Upper} represents the total time taken for decision making when the \emph{challenge\_response} packet is generated, while \emph{Lower} represents the time taken for decision making during the initial scrutiny phase. 
\begin{table}[h]
\begin{center}
\caption {Time taken (in seconds) to detect traffic conditions with varying RSU and POC distances}
\resizebox{\linewidth}{!}{%
\begin{tabular}{|l|*{7}{c|}}\hline
\backslashbox{RSU Distance}{POC Distance}
&\makebox[1em]{10}&\makebox[1em]{15}&\makebox[1em]{20}
&\makebox[1em]{25}&\makebox[1em]{30}&\makebox[1em]{35}&\makebox[1em]{40}\\\hline
2500 Lower & 21 & 53 & 84 & 104.0 & 190.3 & 256 & 326.8 \\
\hline
2500 Upper & 79 & 111 & 141 & 162.0 & 248.3 & 313 & 384.3 \\
\hline
2000 Lower & 23 & 54 & 84.33 & 102 & 193.67 & 254 & 327 \\
\hline
2000 Upper & 69 & 101 & 130.33 & 147 & 239.67 & 301 & 375 \\
\hline
1500 Lower & 22 & 55 & 82 & 101.7 & 192.5 & 257 & 327 \\
\hline
1500 Upper & 59 & 92 & 120 & 138.7 & 230.5 & 294.85 & 366 \\
\hline
1000 Lower & 24 & 55 & 82.5 & 103.2 & 194.2 & 257 & 326 \\
\hline
1000 Upper & 53 & 83 & 111.5 & 132.2 & 222.2 & 286 & 354 \\
\hline
\end{tabular}}
\end{center}
\end{table}
%\vspace*{-1 em}
\begin{figure}[h]
    \centering
 %    \resizebox {0.45\textwidth} {!} {
    \begin{tikzpicture}
\begin{axis}[
    footnotesize,
      title={},
      legend style={nodes={scale=0.5, transform shape}}, 
       % legend image post style={mark=*},
    xlabel={POC distance},
    ylabel={Time in seconds},
    xmin=10, xmax=40,
    ymin=50, ymax=400,
    xtick={10,15,20,25,30,35,40},
    ytick={50,100,150,200,250,300,350,400},
    legend pos= outer north east,
    ymajorgrids=true,
    legend entries={2500 meters Upper,
                    2500 meters Lower,
                    2000 meters Upper,
                    2000 meters Lower,
                    1500 meters Upper,
                    1500 meters Lower,
                    1000 meters Upper,
                    1000 meters Upper},
    grid style=dashed,
]

\addplot[
    color=green,
    mark=triangle,
    ]
    coordinates {
    (10,79)(15,111)(20,141)(25,162)(30,248.3)(35,313)(40,384.3)
    }; %\addlegendentry[text width=45pt, text depth=]{Proposed Model}

\addplot[
    color=blue,
    mark=square,
    ]
    coordinates {
    (10,23)(15,54)(20,84.33)(25,102)(30,193.67)(35,254)(40,327)
    }; %\addlegendentry[text width=45pt, text depth=]{Proposed Model}
 
\addplot[
    color=red,
    mark=x,
    ]
    coordinates {
    (10,69)(15,101)(20,130.33)(25,147)(30,239.67)(35,301)(40,375)
    }; %\addlegendentry[text width=45pt, text depth=]{Majority Voting}

\addplot[
    color=black,
    mark=o,
    ]
    coordinates {
    (10,22)(15,55)(20,82)(25,101.7)(30,192.5)(35,257)(40,327)
    }; %\addlegendentry[text width=45pt, text depth=]{Peer Authentication}

\addplot[
    color=orange,
    mark=oplus,
    ]
    coordinates {
    (10,59)(15,92)(20,120)(25,138.7)(30,230.5)(35,294.85)(40,366)
    }; %\addlegendentry[text width=45pt, text depth=]{Proposed Model}

\addplot[
    color=violet,
    mark=asterisk,
    ]
    coordinates {
    (10,24)(15,55)(20,82.5)(25,103.2)(30,194.2)(35,257)(40,326)
    }; %\addlegendentry[text width=45pt, text depth=]{Proposed Model}

\addplot[
    color=green,
    mark=star,
    ]
    coordinates {
    (10,53)(15,83)(20,111.5)(25,132.2)(30,222.2)(35,286)(40,354)
    }; %\addlegendentry[text width=45pt, text depth=]{Proposed Model}

\addplot[
    color=black,
    mark=oplus,
    ]
    coordinates {
    (10,21)(15,53)(20,84)(25,104)(30,190.3)(35,256)(40,326.8)
    }; %\addlegendentry[text width=45pt, text depth=]{Proposed Model}
\end{axis}
\end{tikzpicture}
\caption{Broadcast comparison of 100 vehicles with varying thresholds}
    \label{fig:my_label}
\end{figure}
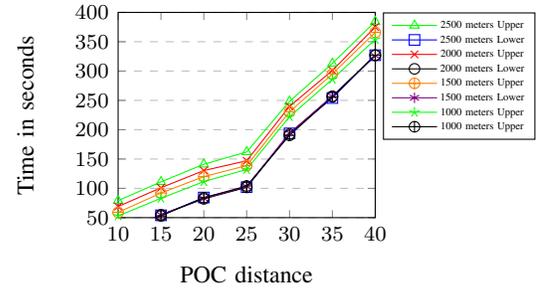
Table III also reveals the limitation of the proposed model. It can be seen that the time taken to detect the traffic condition is not in real time. This is because the amount of time lapsed in sending the \emph{data\_packet}s and detecting the traffic condition is not real time. Thus, this restricts the application of the model in scenarios that do not require real time decisions. This is justified by the fact that the ROI is usually very large, and the main purpose of the proposed model is mainly to allow the vehicles in one ROI to know the traffic scenario of another ROI that it wants to enter. However, the proposed model cannot be applied in autonomous vehicles that require decisions in negligible time, such as turning the steering wheel or braking on the road. Based on the values in Table III, the \emph{detection\_probability} (defined as the ratio of the number of observations where the road condition is detected within a time limit to the total number of observations) is shown in Figure 8. It is to be noted that as the time increases, \emph{detection\_probability} increases. This is because with more time, conflicting vehicles at a higher \emph{POC} distance are also considered that increase the accuracy of the model. Hence, the probability of finding a \emph{POC} and detecting more scenarios increases as more \emph{POC} distances are covered, and eventually, it leads to an increase in the \emph{detection\_probability} at the cost of time. For example, in Table III, the number of observations with the time limit below 50 seconds is \emph{4} (21,23,22,24). This means that 4 out of 56 observations (50\% of the observations with \emph{POC} distance 10) is detected within \emph{50} seconds, thereby covering approximately \emph{7\%} of the total observations that detect the road condition with a \emph{detection\_probability} of \emph{0.07}. However in Table III, the number of observations detected below \emph{100} seconds is \emph{16}, which covers every observation with the \emph{POC} distance of 10, \emph{80\%} of the observations with the \emph{POC} distance of 15, and about \emph{50\%} of the observations with the \emph{POC} distance of 20. This covers almost \emph{28.6\%} of the total observations in Table III (\emph{detection\_probability} is \emph{0.28}) as compared to \emph{7\%}, when the time limit was below 50 seconds.

$detection\_probability=\frac{total\_detections\_under\_certain\_time}{total\_possible\_ detections}$
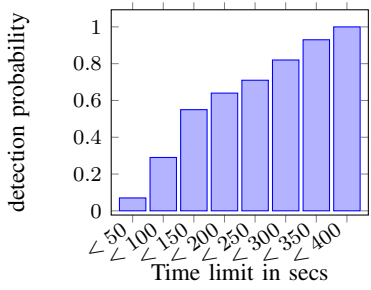
\begin{figure}
    \centering
\begin{tikzpicture}
\begin{axis}[
ybar,
footnotesize,
ylabel=detection probability,
	xlabel = Time limit in secs,
	symbolic x coords={$<50$,$<100$,$<150$,$<200$,$<250$,$<300$,$<350$,$<400$},
	x tick label style={rotate=35,anchor=east}
	]
\addplot coordinates
{($<50$,0.07) ($<100$,0.29)
		 ($<150$,0.55) ($<200$,0.64)($<250$,0.71)($<300$,0.82)
		 ($<350$,0.93) ($<400$,1.0)};
\end{axis}
\end{tikzpicture}
    \caption{Detection probability with respect to time}
    \label{fig:my_label}
\end{figure}
\section{Security and Privacy Analysis of Our Model}
\thispagestyle{empty}
\subsection{Security Analysis}
\subsubsection{Message and GPS information spoofing attack}
In our proposed model, every vehicle records an \emph{event} and generates a \emph{data\_packet}. In order to disrupt the decision making process, it can report an inaccurate event, i.e., report a congestion when the road is non-congested. For the attack to be successful, every vehicle at the ROI must spoof the event, which would incapacitate the edge server to detect any \emph{POC}. However, this contradicts our assumption that at least one vehicle must be non-malicious. Furthermore, an attacker may spoof the \emph{GPS} information, i.e., sent manipulated \emph{vel\textsubscript{id}} and \emph{GPS\textsubscript{id}} to the edge server. However, in the proposed heuristic, the edge server leverages \emph{vel\textsubscript{id}} and \emph{GPS\textsubscript{id}} to generate the \emph{challenge\_packet} to filter malicious vehicles within the ROI. Thus, our proposed model is secure against message and GPS information spoofing attack.
\subsubsection{Masquerading, Collusion and Sybil attack}
In the proposed model, every vehicle digitally signs its \emph{V\textsubscript{id}} using \emph{veh\_private\textsubscript{i}} generated using \emph{EGDSS} to produce \emph{ds(V\textsubscript{id})}. An attacker can forge the signature if it can compute \emph{veh\_private\textsubscript{i}} of the attacked \emph{V\textsubscript{id}} or by performing \emph{hash collision attack}. However, SHA-3 \cite{dworkin2015sha} is secured against a collision attack, preimage \cite{aoki2008preimage} and second preimage attack\cite{yu2005second} with the security strength varying from \emph{112-256} bits for collision and \emph{224-512} bits for preimage and second preimage attack.
An attacker may also perform collusion or a sybil attack to disrupt the decision making process. However, for the attack to be successful, every \emph{V\textsubscript{id}}s within the ROI must be malicious. This contradicts our assumption of having one non-malicious vehicle within the ROI. Thus, the proposed model is secured against masquerading, collusion and sybil attacks.
\subsubsection{Message Integrity Attack}
In the proposed model, every vehicle sends ${encrypted\_data\_packet} = <\tau,data\_packet`>$ to an edge server. If an attacker wants to violate the integrity of a \emph{data\_packet} generated by a \emph{V\textsubscript{id}}, he/she must acquire \emph{G\textsubscript{private}} stored only at the edge server, i.e. an attacker has to compute \emph{p} and \emph{q} used by the edger server to generate \emph{G\textsubscript{private}}. However, this violates the discrete logarithm problem \cite{mccurley1990discrete}. An attacker may try to compute \emph{Key\textsubscript{i}} generated using AES 128 to obtain \emph{data\_packet}. However, supercomputer will take around \emph{1 billion years} to brute-force the key \cite{arora2012secure} while the \emph{biclique attack} \cite{bogdanov2011biclique} requires a computational complexity of \emph{2\textsuperscript{126.1}}, which is highly unlikely to break in real-time. Thus, our proposed model is resilient against message integrity attack.   
\subsubsection{DoS attack}
Let us assume that the number of vehicles within the ROI is \emph{V\textsubscript{num}}.The various combinations of malicious vehicles not sending the packet to the edge server, defined as C(DoS), is given by:

$\hspace*{20mm}C(DoS)=\sum_{i=1}^{V\textsubscript{num}} {{V\textsubscript{num}}\choose{i}}$

C(DoS) represents the different number, ranging from 0 to V\textsubscript{num}, of vehicles that can refrain from sending the information to the edge server via the RSU. However, \emph{DoS} is possible only when every vehicle within the ROI drops their \emph{data\_packet}s. Thus, the probability of the \emph{DoS} attack being successful, defined as \emph{P(DoS)}, contradicts our assumption that every non-malicious vehicle sends their packets, at least one packet will be received by the edge server, and also that one non-malicious vehicle should be present within the ROI. Thus, the proposed model prevents \emph{DoS} attacks. 
$\hspace*{20mm}P(DoS)=\frac{1}{C(DoS)}$
\subsection{Privacy Analysis}
\subsubsection{Conditional Privacy Preservation}
In our proposed model, every vehicle generates \emph{enc\_id} by encrypting its \emph{V\textsubscript{id}} with \emph{G\textsubscript{public}}. If an attacker attempts to track the  \emph{V\textsubscript{id}} of a vehicle, it has to compute \emph{G\textsubscript{private}} using the associated value of prime numbers, \emph{p} and \emph{q}, possessed only by  the edge server. In order to breach the privacy of the \emph{data\_packet} of a vehicle, it has to obtain the \emph{Key\textsubscript{i}}, generated using AES 128, used by a \emph{V\textsubscript{id}} to encrypt the \emph{data\_packet}. Even under such a scenario, the attacker has to compute \emph{G\textsubscript{private}} as \emph{Key\textsubscript{i}} is encrypted using \emph{G\textsubscript{public}}. Thus, the proposed model guarantees anonymity and unlinkability of a vehicle. However, the edge server filters malicious vehicles using the \emph{DSG}. Thereafter, it is stored in the centralized server for future reference. Thus, the proposed model also guarantees conditional privacy, and only reveals the identity of the malicious vehicles when it detects a conflict. 

$P(DoS)=\frac{1}{C(DoS)}$
\section{Conclusion}
In this paper, we proposed a privacy preserving secure edge cloud-assisted traffic monitoring system for VANETs that provides accurate traffic-related information %within the ROI
. The proposed model is resilient against privacy attacks and unauthorized tracking, and is secured against collusion, masquerading, ballot stuffing, and bad mouthing attacks. We used \emph{DSG} and the \emph{challenge-response} strategy to filter malicious responses, and to determine accurate traffic-related information with fewer number of broadcasts per vehicle compared to the peer authentication model. Even though the number of broadcasts per vehicle required for the proposed model is higher than the majority voting model and the reputation based model, the proposed model has a higher \emph{detection\_accuracy} when the number of malicious vehicles forms the majority within the ROI. This means that unlike the majority voting and the reputation-based model, the proposed model filters malicious vehicles and accurately detects the traffic condition %of the ROI 
under the influence of at least one non-malicious vehicle. In future, we plan to extend our work using untrusted RSUs distributed sparsely throughout the ROI, such as in semi-urban and rural areas, which delays the information exchange between the vehicle and the RSU. We will also design an intrusion detection system (IDS) for the in-vehicle network that can detect any fabricated information injected within the a vehicle without any V2X communications.\thispagestyle{empty}
\bibliographystyle{unsrt}
%\vspace*{-1em}
\bibliography{ref}
\thispagestyle{empty}

\end{document}